\documentclass[12pt]{iopart}

\usepackage{hyperref}
 
\usepackage[dvips]{graphicx}
 
\def\v#1{{\bf#1}}
\def\be{\begin{equation}}
\def\ee{\end{equation}}
\def\bea{\begin{eqnarray}}
\def\eea{\end{eqnarray}}

\newcommand{\bfalpha}{\mbox{\boldmath$\alpha$\unboldmath}}

\def\ie{{\it i.e.\,}}

\def\<{\langle}
\def\>{\rangle}
 
\begin{document}

\title{Relativistic echo dynamics and the stability of a beam of Landau electrons}
 
 
\author{E Sadurn\'i $^1$ and T H Seligman $^{1,2}$ }
 
\address{$^1$Instituto de Ciencias F\'isicas,
Universidad Nacional Aut\'onoma de M\'exico,
Cuernavaca, Morelos, M\'exico.}
 
 
\address{$^2$Centro Internacional de Ciencias,
 Cuernavaca, Morelos, M\'exico.}
 
 
\ead{sadurni@fis.unam.mx, seligman@fis.unam.mx}

 
\begin{abstract}

We extend the concepts of echo dynamics and fidelity
decay to relativistic quantum mechanics, specifically
in the context of Klein-Gordon and Dirac equations
under external electromagnetic fields. In both cases
we define similar expressions for the fidelity
amplitude
under perturbations of these fields, and a
covariant version of the echo operator.
Transformation properties under the Lorentz group are
established. An alternate expression for fidelity is
given in the Dirac case in terms of a 4-current. As an
application we study a beam of Landau electrons perturbed by
field
inhomogeneities.

\end{abstract}
 
 
 
\maketitle

Echo dynamics and fidelity decay have received considerable attention in recent years
\cite{review}. Their importance is underlined by the fact that fidelity is a standard benchmark
in quantum information \cite{nielsen}. 
The relevance of these concepts is
centered on non-relativistic quantum mechanics. 
It is rather surprising, that relativistic
problems including quantum field theoretical ones have
not been considered in this context. 
 
In this letter we shall formulate echo dynamics and
fidelity decay concentrating
on the Dirac and Klein-Gordon equations, as the
simplest cases. Field theoretical considerations or
more complicated equations for particles with higher spins will be left
for future work, though we hope that the concepts
developed will readily generalize.
On the other hand the inclusion of external fields is
essential, as these will be the 
source of any reasonable perturbation. These
fields will be assumed to result 
from a Lorentz covariant theory and couple
minimally to the particle, although this last
requirement can be relaxed in view of the intrest in
systems such as the Dirac oscillator \cite{DO}. We
shall limit our considerations both for simplicity,
and because
of the predominant practical importance  to
electromagnetic fields with their Abelian
gauge structure. Let us stress that even if Dirac theory is thought to
contain many particles already in first quantization,
here we deal with the Dirac equation of one particle
in the sense that just one set of space-time
coordinates appears and we have to worry about
evolution in one time only. In the light of past work \cite{review}, time scales
of fidelity decay are of utmost importance when
interpreting results. Here we would like to point out
that relativity introduces a new time scale in terms
of
the speed of light, whose physical meaning is strongly related to the intensities of perturbations. This scale is given by $\hbar / m c^2$, $m$ being the rest mass. Assuming that $\lambda$ actually characterises the strength of the perturbation in energy units, we can compare to the standard timescale of fidelity decay $\hbar/\lambda$.

It is known \cite{fw} that relativistic effects
appear when $\lambda > mc^2$, but also when the
elapsed time $t$ is long enough for the relativistic
propagator to be significantly different from a delta
function and therefore from its non relativistic
counterpart. This means that for $t > \hbar/mc^2 >
\hbar/\lambda $,
 quasi relativistic expansions
are not useful, motivating thus a full relativistic
treatment. Let us
 mention that our approach also includes the so called
{\it echo kernel},
 which generalizes the concept of the echo operator
\cite{review}.
 Its importance rests in the fact that it allows a
perturbative treatment
 for arbitrarily long times, if the perturbation is
weak enough, while the evolution operator does not. 
 
We shall first discuss the problem for the
Klein-Gordon equation and then proceed to 
the Dirac case. For the latter we can give an
interesting alternate, though equivalent, formulation
in terms of currents, which will be then used to
discuss the perturbation 
of a Landau electron by field inhomogeneities. We
close with some comments on 
possible generalizations. Let us consider two physical systems described by the
Klein-Gordon
 equation with interaction. The
systems will be
 considered to be minimally coupled to Lorentz
covariant 4-vector potentials.
 One of the systems will be subject to a
peturbation giving
 rise to a modified evolution of the wave function.
We wish to give an expression for the corresponding fidelity.
For simplicity we
 use units $\hbar = c = e = 1$ where $e$ is the charge
of the electron and
$c$ the speed of light.
 All space integrations will be taken over $R^3$
unless indicated
 otherwise. The wave equation which describes the
perturbed system contains the
 unperturbed one as a special case. It is given by
\bea
\hat O_{KG}(\epsilon) \phi(\epsilon) \equiv \left[
D_{\mu}(\epsilon)
 D^{\mu}(\epsilon) + m^2 \right] \phi(\epsilon) = 0 
\label{1}
\eea
where
\bea
D^{\mu}(\epsilon) = \partial^{\mu} + i B^{\mu} + i
\epsilon A^{\mu},
 \qquad D^{\mu} = D^{\mu}(0)
\label{3}
\eea
are the covariant derivatives with a ``background''
field
 $B^{\mu}$ and a perturbation $A^{\mu}$ modulated by
$\epsilon$. By means of
 the appropriate definition of the inner product in
Klein-Gordon theory
 \cite{dewitt}, namely
\bea
\< \phi_1, t | \phi_2, t \> = i \int d^3x  \left[
\phi^{*}_1(x,t)
  \partial_0 \phi_2(x,t) - (1 \leftrightarrow 2)^{*}
\right],
\label{4}
\eea
we can write the fidelity amplitude for a
given initial
 condition $\phi(x)$. In fact, defining $t$ as the 0
component of both (final)
 events $x_{\mu}$ and $x'_{\mu}$, the fidelity
amplitude $f$ after a
 time $t$ is given by
\bea
f(t) \equiv \< \phi(\epsilon), t | \phi, t \> = \int
d^3x' \int d^3x
 \phi^{*}(x)\phi(x')M(x_{\mu},x'_{\mu}),
\label{5}
\eea
where an \textit{echo kernel} $M$ has been defined in
terms of the
 Lorentz invariant propagator \cite{kaku, ryder} (or Feynman
 propagator) of equation (\ref{1}). This propagator
satisfies the equation 
\bea
\hat O_{KG}(\epsilon)
\Delta_{\epsilon}(x_{\mu},x'_{\mu}) = i
 \delta^4(x_{\mu}-x'_{\mu}), \qquad \Delta =
\Delta_{\epsilon=0}
\label{6}
\eea
and the echo kernel is found to be
\bea
M(x_{\mu},x'_{\mu}) = \int d^3x'' \left[
 \Delta^{*}_{\epsilon}(x''_{\mu},x_{\mu})
\frac{\partial}{\partial t}
\Delta(x''_{\mu},x'_{\mu})-
 \Delta(x''_{\mu},x'_{\mu}) \frac{\partial}{\partial
t}
 \Delta^{*}_{\epsilon}(x''_{\mu},x_{\mu}) \right].
\label{7}
\eea
At this point a word must be said about the existence
of an echo
 operator related to kernel (\ref{7}). Despite the
presence of $M$ in
 (\ref{5}), the lack of a standard hamiltonian
formulation of the Klein-Gordon
 equation \cite{find, dewitt} (\ie, one with a
hermitian Hamilton
 operator) forbids us to relate (\ref{7}) in a direct
way to a unitary
 evolution operator as in Schroedinger theory. 
 
The Lorentz transformations for the Klein-Gordon
fidelity amplitude and
 echo kernel are most simply derived by recalling the
Lorentz
 invariance of the Feynman propagator
$\Delta_{\epsilon}$ and the transformation
 properties of the volume element. With this, through (\ref{5}), it is
 easily shown that the effect of a boost in an arbitrary
direction with
 associated Lorentz factor $\gamma$ results in
\bea
f(t)= f(\gamma t'), \qquad \gamma= 1/\sqrt{1-v^2}
\label{7.1}
\eea
where $v$ is the boost velocity and $t'$ is the
observed time in the
 boosted frame. Thus the form of the
amplitude is unchanged,
 while transformation acts exclusively on its argument
as a time
 difference.
 
Next we discuss the fidelity amplitude and echo
operator for the Dirac
 equation. Consider a system with background
field and perturbation as above. It is now described by the
equation
\bea
\hat D(\epsilon) \psi(\epsilon) \equiv \left[ -i
\gamma^{\mu}
 D_{\mu}(\epsilon) + m \right]\psi(\epsilon) = 0
\label{8}
\eea
where $D_{\mu}(\epsilon)$ is given by (\ref{3}) and
$\gamma^{\mu}$ are
 Dirac matrices \cite{dirac}. Since this theory allows
a dynamical
 description in terms of a Hamiltonian, the concepts
of fidelity and echo
 operator need little or no modification from the
usual ones. The fidelity amplitude is thus given
by
\bea
\< \psi(\epsilon), t | \psi, t \> = \< \psi(\epsilon)
|
 U^{-1}_t(\epsilon) U_t  | \psi \> = \< \psi(\epsilon)
| M_t  | \psi \>
\label{10}
\eea
where $|\psi \> = |\psi,0 \>$ is our initial condition
and
 $U(\epsilon)$ the unitary evolution
operator. When $H_D$ is set as the Dirac Hamiltonian
associated to (\ref{8}), $U$ satisfies
\bea
H_D U(\epsilon) = -i \frac{\partial}{\partial t}
U(\epsilon), \qquad U
 = U(\epsilon=0).
\label{11}
\eea
Since a Schroedinger-like equation can be reached from
(\ref{8}) by
 factoring $\gamma^0$ from the Dirac operator $\hat
D(\epsilon)$, we may
 prefer to handle evolution operators and propagators
directly
 related to the Lorentz invariant equation (\ref{8})
rather than
 (\ref{11}). Thus, defining
\bea
U'= U \gamma^0, \qquad M'_t= (U'_t(\epsilon))^{-1}
U'_t \gamma^0 =
 \gamma^0 M_t
\label{12}
\eea
we are led to the Lorentz invariant propagator
\bea
K'_{\epsilon}(x_{\mu},x'_{\mu}) = \< x |
U'_t(\epsilon) | x' \>
\label{13}
\eea
and the echo kernel
\bea
M'(x_{\mu},x'_{\mu}) = \< x | M'_t | x' \>,
\label{13.1}
\eea
where $t$ is the elapsed time (or time
difference) between events
 $x_{\mu}$ and $x'_{\mu}$. The fidelity amplitude is
written in terms
 of the Dirac inner product as
\bea
f(t)= \< \psi(\epsilon), t | \psi, t \> = \int d^3x
 \psi^{\dagger}_{\epsilon}(x,t) \psi(x,t).
\label{14}
\eea
It has the correct Lorenz transformation property
(\ref{7.1}) under
application of a boost with the corresponding $\gamma$
factor.
 
The Dirac equation allows another and more interesting
approach to
 fidelity by means of a 4-vector which resembles a
Dirac current. Consider a
 bilinear form in the wave functions given by
\bea
j_{\mu}(\epsilon) = \bar \psi(\epsilon) \gamma_{\mu}
\psi, \qquad \bar
 \psi = \gamma^0 \psi^{\dagger}.
\label{16}
\eea
The fidelity amplitude is obtained by integrating the
$0$
 component of (\ref{16}) 
\bea
f(t)= \int d^3x j_0 (\epsilon).
\label{17}
\eea
This displays clearly, that for $\epsilon=0$, the
fidelity amplitude
 becomes unity by conservation of probabilty. In fact,
when the
 perturbation is present, the bilinear form (\ref{16})
obeys  a
 continuity-like equation
\bea
\partial^{\mu} j_{\mu}(\epsilon) = i \epsilon A^{\mu}
j_{\mu}(\epsilon)
\label{18}
\eea
where the Abelian character of $A$ has been used.
Equation (\ref{18})
 indicates that the conservation law is now corrected
by a source given
 exclusively in terms of the perturbation applied to
the system.
 Following this line of reasoning, (\ref{18}) can be
integrated over the space
 variables of some inertial observer to yield
\bea
\frac{df(t)}{dt} = \int_S \psi^{\dagger}(\epsilon) d
\v s \cdot
 \bfalpha \psi -i \epsilon \int_V d^3x
A^{\mu}j_{\mu}(\epsilon)
\label{19}
\eea
with $V$ the volume of integration and $S$ its
boundary surface. The
 appearence of a boundary term can be related to
fidelity loss when
 dealing with finite portions of space-time or with
special boundary
 conditions for which wave functions do not vanish at infinity
(e.g. unbound states). Its contribution is present independently
of $\epsilon$. Considering an infinite volume of integration $V$
and a vanishing
 boundary term (bound states) we further simplify
 (\ref{19}) using an expansion in $\epsilon$ to
lowest order. 
In this  approximation 
fidelity amplitude and fidelity respectively
 obey the differential equations
\bea
\frac{df(t)}{dt} = -i \epsilon \int d^3x
A^{\mu}j_{\mu} = -i \epsilon
 \< \gamma^0 \gamma^{\mu} A_{\mu}  \>_I,  f(0)=1,
\label{20}
\eea
\bea
\frac{dF(t)}{dt} = -2 \epsilon Re \left[ i f(t) \<
\gamma^0
 \gamma^{\mu} A_{\mu}  \>_I \right],  F(t)=|f(t)|^2
\label{21}
\eea
where $\< \cdot \>_I$ is the average 
with respect to $\psi$ in the
 interaction picture. These two simple equations can
be used to discuss some
 application.
 
We study Landau electrons 
perturbed by an
 inhomogeneous magnetic field. The system is described by the Dirac equation (\ref{8}) with a 
homogeneous background field
 which, for some inertial observer, acquires the form
$\v B = \v H \times \v
 r$, $B_0=0$. \ie  a constant magnetic field of
intensity $H$. Here $\v
 r$ is the space part of $x_{\mu}$ for such an
inertial observer. In this
 frame, when $t=x_0=0$ we switch on an additional
field given by the
 perturbative static potential $\epsilon A$ whose
components are bounded
 such that the perturbation of strength  $\epsilon$ is
meaningful at all space-time points.
 If the system is stationary at $t<0$, it will change to a non-stationary state
at $t>0$ and we
 would like to know how fidelity evolves.
 
The system at negative
times is well known
 to be integrable. Its energy eigenstates are
infinitely degenerate \cite{landau, olandau}. A number of
general results for fidelity amplitudes in the case of perturbed integrable
systems have been established \cite{review}. Some interesting features can be exploited
here. The results for this setup will be valid not only for one electron
propagating freely in
 the direction parallel to the field, but also for a
beam of electrons
 under certain (physical) considerations. Although the
treatment we have
 given does not contemplate many Dirac particles in
the fashion of an
 extended configuration space \cite{moshsadur, moshnikitin}, we
 may consider the evolution to be described by a single time. The electromagnetic interactions
between the constituents of
 the particle beam will be neglected, implying
the absence of
 quantum field effects. This requires a wavelength
in the propagating
 direction greater than Compton's as well as a low
particle density in the beam.
 
With these physical restrictions and taking advantage
of the infinite
 degeneracy of the unperturbed system, it is possible
to accomodate an
 arbitray number of particles in a stationary state without
violating the Pauli principle. We find that (\ref{21})
(or equivalent
 expressions (12), (13) of \cite{review}) yields a
decay of fidelity
 dominated by a quadratic term in $t$ and with
negligible oscillatory terms as we shall see in detail. When computing correlations in (12) and (13) from \cite{review}, the relevant term is given by
\bea
\int_{0}^{t} dt' \int_{0}^{t} dt'' \<i| \tilde V(t') \tilde V(t'')  |j \> \nonumber \\ = 4 \sum_{j} | \<i| \gamma_0 \gamma_{\mu} A^{\mu}|j \>|^2 \frac{\sin^2{ \left[(E_i-E_j)t/2 \right] }}{(E_j-E_i)^2} 
\label{21detail}
\eea
where $i$ and $j$ indicate all quantum numbers including momentum, $E_i, E_j$ being the corresponding energies. All degenerate levels will contribute to the dominant term of the sum in (\ref{21detail}) even though $ \<i| \gamma_0 \gamma_{\mu} A^{\mu} |i \> = 0 $, which can be assumed without loss of generality \cite{reviewcomment}. The correlation can be approximated by
\bea 
C(H,k) t^2 \equiv \left( \sum_{j: E_j = E_i} |<i|\gamma_0 \gamma_{\mu} A^{\mu}|j>|^2 \right) t^2 
\label{22detail}
\eea 
where $k$ is the $z-$momentum of the particle at $t<0$. The $H$ dependence enters through the unperturbed wave functions. We shall denote by $C(H)$ the coefficient in (\ref{22detail}) when $k=0$. When dealing with a beam of particles we may replace 
\bea
C(H,k) = \sum_{n: E_n=E_i} w_n \left( \sum_{j: E_j = E_i} | \<n|\gamma_0 \gamma_{\mu} A^{\mu}|j \>|^2 \right) 
\label{23detail}
\eea
where $w_n$ are statistical weights for states in the beam with energy $E_i$. Result (\ref{23detail}) is not surprising from the point of view of
integrability
 of the unperturbed system. Nevertheless it is
remarkable in a theory of many fermions due to the
infinite degeneracy of
 levels. 

The result mentioned above can also be derived in
the non-relativistic
 version of this problem. Lorentz
transformations enter the game if we recognize that the momentum of the beam $k$
is one of the
 control parameters (the others being $H$ and
$\epsilon$) and that plane
 waves of arbitrary momentum can be obtained by
applying boosts in the
 direction of propagation. From property (\ref{7.1}),
an increase of momentum
 $k$ results in a Lorentz factor
$1/\sqrt{1-v(k)^2}$, thus changing the rate of
fidelity decay. 
 These considerations lead finally to a decay of 
fidelity amplitude and fidelity given by
\bea
f(t) \sim 1-\epsilon^2 C(H) \left( 1-(v(k))^2
\right) t^2/2,
\label{21.1}
\eea
 
\bea
F(t) \sim 1 - \epsilon^2 C(H) \left( 1-(v(k))^2
\right) t^2.
\label{22}
\eea
The variable $t$ 
denotes the elapsed time
as measured in the frame indicated above and the
momentum dependent
 velocity is
\bea
v(k)= \frac{k}{\sqrt{k^2+m^2}}.
\label{23}
\eea
Restoring our physical constants, the Compton
wavelength restriction
 can be put as
\bea
k < 2mc, \quad v < 2c/\sqrt{5}, \quad
\sqrt{1-(v(k)/c)^2} > 1/\sqrt{5}
\label{24}
\eea 
resulting in a lower bound for the magnitude of the
decay rate. The upper bound for the momentum implies a
maximum relativistic time
 delay. Allowing pair creation would slow down
considerably the decay
 of fidelity, as long as $C(H)$ remains under control.
 
Summarizing, 
we have proposed a consistent formulation of
echo dynamics and fidelity decay 
for the simplest relativistic wave equations. 
The perturbative regime, which has far reaching
implications and applications
in the non-relativistic case, yields an interesting
behavior.
We apply our formulation to a beam of Landau electrons
propagating along a homogeneous magnetic field perturbed by bounded inhomogeneities.

Our treatment was limited to the discussion of a
single particle with the corresponing single time
line.
Yet for systems of many particles, generalizations of
the Dirac equation have been 
proposed; these involve in principle many {\it times}
but a possibility for a single
  time evolution has been devised \cite{moshsadur, omoshsadur}. 
Ultimately, composite relativistic systems can be
treated along the lines  we 
proposed here, since the Lorentz structure of
evolution
equations is not altered by
 extending configuration space (multiple space-time
coordinates, one set for each
 particle)
We also restricted our considerations to special
relativity, but many of the results presented in this
work can be generalized to curved spaces or
``background'' metrics by merely introducing covariant
derivatives with the appropriate connections
\cite{gr-con} and the corresponding transformation of
spinors for the Dirac case \cite{gr-spin}.
Considering these facts our proposition for a
relativistic  formulation of
fidelity and echo dynamics seems adequate 
or at least readily adaptable to the
treatment of more general 
situations then the one we actually discuss. The issue of accelerated observers will be the
subject of future work.
 
\ack
 
We are greatful to M. Moshinsky for a life of
discussions and
we aknowledge support under the Grants PAPIIT IN112507 and
CONACyT 57334.

\end{document}